\documentclass[11pt]{amsart}
\usepackage{amsmath,amsfonts,amssymb,amsthm,amscd}
\newtheorem{thm}{Theorem}

\newtheorem{prop}[thm]{Proposition}

\newtheorem{example}[thm]{Example}

\def\res{\mathop{\rm res}\limits}
\begin{document}

\title[Rational and Trigonometric solutions of the WDVV equations]{A note on the relationship between
rational and trigonometric solutions of the WDVV equations}

\author{Andrew Riley$^{(1)}$ and Ian A. B. Strachan$^{(2)}$}

\date{}
\address[1]{Department of Mathematics\\ University of Hull\\
Hull HU6 7RX\\ U.K.}

\email{a.riley@math.hull.ac.uk}

\address[2]{Department of Mathematics\\ University of Glasgow\\ Glasgow G12 8QQ\\ U.K.}

\email{i.strachan@maths.gla.ac.uk}


\begin{abstract}
Legendre transformations provide a natural symmetry on the space of solutions to the WDVV equations,
and more specifically, between different Frobenius manifolds. In this paper a twisted Legendre
transformation is constructed between solutions which define the corresponding dual Frobenius manifolds.
As an application it is shown that certain trigonometric and rational solutions of the WDVV equations are
related by such a twisted Legendre transform.
\end{abstract}

\maketitle

\section{Introduction}

The Witten-Dijkgraaf-Verlinde-Verlinde (or WDVV) equations of associativity
\[
\frac{\partial^3F}{\partial t^\alpha \partial t^\beta \partial t^\lambda}\eta^{\lambda\mu}
\frac{\partial^3F}{\partial t^\mu \partial t^\gamma \partial t^\delta}-
\frac{\partial^3F}{\partial t^\delta \partial t^\beta \partial t^\lambda}\eta^{\lambda\mu}
\frac{\partial^3F}{\partial t^\mu \partial t^\gamma \partial t^\alpha}=0
\,,\quad \alpha\,,\beta\,,\gamma\,,\delta=1\,\ldots\,,n
\]
have been much studied from a variety of different points of view, amongst them,
topological quantum field theories, Seiberg-Witten theory, singularity theory and integrable systems.
Geometrically, a solution defines a multiplication $\circ: TM \times TM \rightarrow TM$ of vector fields, i.e.
\begin{eqnarray*}
\partial_{t^\alpha} \circ \partial_{t^\beta} & = &\left(
\frac{\partial^3F}{\partial t^\alpha \partial t^\beta \partial t^\sigma} \eta^{\sigma \gamma}\right)
\partial_{t^\gamma}\,,\\
& := & c_{\alpha\beta}^\gamma(t)\,\partial_{t^\gamma}\,.
\end{eqnarray*}
The metric $\eta\,,$ used to raise and lower indices, is flat and the coordinates
$\{ t^\alpha \}$ are flat coordinates, i.e. the components
of the metric is this coordinate system are constants. It will be convenient in what follows to denote the metric
as $<,>\,.$

A symmetry of the WDVV equations are transformations
\begin{eqnarray*}
t^\alpha & \longmapsto & {\hat t}^\alpha\,, \\
\eta_{\alpha\beta} & \longmapsto & {\hat\eta}_{\alpha\beta}\,, \\
F & \longmapsto & {\hat F}
\end{eqnarray*}
that preserve the equations. These were studied in \cite{dubrovin1} and in particular
Legendre-type transformations $S_\kappa$ ($\kappa=1\,,\ldots\,,n)$ act as follows:
\begin{eqnarray*}
{\hat t}_\alpha & = & \partial_{t^\alpha} \partial_{t^\kappa} F(t)\,,
\qquad ({\rm N.B.}\quad{\hat t}_\alpha=\eta_{\alpha\beta}t^\beta)\\
\frac{\partial{\hat F}}{\partial{\hat t}^\alpha \partial{\hat t}^\beta} & = &
\frac{\partial{     F}}{\partial{     t}^\alpha \partial{     t}^\beta} \,,\\
{\hat\eta}_{\alpha\beta} & = & \eta_{\alpha\beta}\,.
\end{eqnarray*}
Note that (here we denote $\partial_\kappa$ to be the vector field, which, in the $\{t^\alpha\}$-coordinates, is
the coordinate vector field $\partial_{t^\kappa}$)
\[
\partial_{t^\alpha} = \partial_{\kappa} \circ \partial_{{\hat t}^\alpha}\,.
\]
It follows \cite{dubrovin1} that the new metric $<,>_\kappa$ is related to the original metric by
\[
<a,b>_\kappa = <\partial_\kappa^2,a \circ b>\,,
\]
where $\partial_\kappa^2=\partial_\kappa \circ \partial_\kappa\,,$ and that the variables $\{{\hat t}^\alpha\}$ are
flat coordinates for this new metric.

While such Legendre transformations may be applied to any solutions of the WDVV equations, in this paper
we will concentrate on solutions that define a Frobenius manifold. Such solutions have, by definition,
the quasihomogeneity property
\[
{\mathcal L}_E F = d_F \,F + {\rm quadratic~terms}
\]
where the Euler vector field is linear in the flat-coordinates,
\[
E = \sum_\alpha d_\alpha t^\alpha \partial_{t^\alpha} + \sum_{\alpha|d_\alpha=0} r^\alpha \partial_{t^\alpha}\,.
\]
For the precise definition of a Frobenius manifold see \cite{dubrovin1}\,.

\bigskip

\begin{example}\label{example1} (see \cite{dubrovin1} example B1)
\[
\left\{
\begin{array}{ccc}
F & = & \frac{1}{2} (t^1)^2 t^2 + e^{t^2} \\
E & = & t^1 \partial_{t^1} + 2 \partial_{t^2}
\end{array}
\right\}
\overset{S_2}{\longrightarrow}
\left\{
\begin{array}{ccc}
{\hat F} & = & \frac{1}{2} ({\hat t}^2)^2 {\hat t}^1 + \frac{1}{2} (({\hat t}^1)^2 (\log {\hat t}^1 - \frac{3}{2}) \\
{\hat E} & = & {\hat t}^2 \partial_{{\hat t}^2} + 2 {\hat t}^1\partial_{{\hat t}^1}
\end{array}
\right\}
\]
where
\begin{eqnarray*}
t^1 & = & {\hat t}^2 \,, \\
t^2 & = & \log {\hat t}^1\,.
\end{eqnarray*}

\end{example}

\noindent Note that the transformation $E\rightarrow {\hat E}$ is just a coordinate transformation, so technically
it does not really require the hat as geometrically nothing has changed and so this hat may be dropped.
Similar remarks hold for the
multiplication $\circ$ - this just transforms as a $(2,1)$-tensor under the above
transformation and so does not require a hat.

\bigskip

Recently, it was shown by Dubrovin \cite{dubrovin2} that one may construct a so-called
dual solution to the WDVV equations starting from
a given Frobenius manifold (and hence a specific solution of the WDVV equations). This is constructed in two stages,
from a new multiplication and a new metric.

From the specific Frobenius manifold $M$
one may define a dual multiplication $\star\,: T{\overset{\star}{M}}\rightarrow T{\overset{\star}{M}}$ by
\[
X \star Y = E^{-1} \circ X \circ Y\,,\qquad\forall\,
X\,,Y\in  T{\overset{\star}{M}}\,,
\]
where $E^{-1}$ is the vector field defined by the
equation $E^{-1}\circ E=e$ and $\overset{\star}{M}=M\backslash\Sigma\,,$ where $\Sigma$ is the
(discriminant) submanifold where $E^{-1}$ is undefined and $e$ being the unity vector field on the manifold.
This new multiplication is clearly commutative and associative with the Euler field playing the
role of the unity field for the new multiplication.

On such Frobenius manifold there exists another metric $(,)$ - the intersection form - defined
by the formula
\[
(\omega_1,\omega_2) = i_E(\omega_1 \circ \omega_2)
\]
where $\omega_1\,,\omega_2\in T^*M$ and the metric $\eta$ has been used to extend the multiplication
from the tangent bundle to the cotangent bundle. This metric is also flat and hence there
exists another distinguished coordinate system $\{p^i\,,i=1\,,\ldots\,,n\}$ in which the
components of the intersection form are constant. It follows \cite{dubrovin1} that these
two metrics are related by the formula
\[
(E \circ u,v) = <u,v>
\]
or
\begin{equation}
(u,v) = < E^{-1} \circ u,v>\,,\qquad\forall\,
u\,,v\in  T{\overset{\star}{M}}\,.
\label{metric}
\end{equation}
These new structures are compatible
\[
<X\star Y,Z>=<X,Y\star Z>\,.
\]
From this and various other properties inherited from the original Frobenius structure
one may derive a dual prepotential $F^\star$ satisfying the WDVV equations in the flat coordinated of the intersection
form. Explicitly (here $G_{ij}$ are the compondents on the intersection from in its flat coordinate system):

\medskip

\bigskip

\begin{thm} There exists a function $\overset{\star}{F}(p)$ such that\footnote{In such formulae, Greek indices are
raised and lowered using the metric $\eta$ and Latin indices using the metric $G\,.$}

\[
\frac{\partial^3 {F^\star}(p)}{\partial p^i\partial p^j \partial p^k}=G_{ia}G_{jb}
\frac{\partial t^\gamma}{\partial p^k}
\frac{\partial p^a}{\partial t^\alpha}
\frac{\partial p^b}{\partial t^\beta}c^{\alpha\beta}_\gamma(t)
\]
and which satisfies the WDVV-equations in the $\{p^i\}$ coordinates.

\end{thm}

For the simple two-dimensional examples of Frobenius manifolds given above, it is straightforward to
calculate the flat-coordinates for the two intersection forms and hence to calculate the new multiplication
and prepotentials $F^\star$ and ${\hat F}^\star\,:$

\bigskip

\begin{example}\label{example2} Let $F$ be the prepotential defined in Example 1, with prepotential ${\hat F}$ be constructed
via the Legendre transformation $S_2.$ Let $F^\star$ and ${\hat F}^\star$ be the corresponding dual prepotentials.
Then
\begin{equation}
F^\star=\frac{1}{4} z^1 z^2 (z^1+z^2) - \frac{1}{12} \left( (z^1)^3+(z^2)^3\right) + \frac{1}{2} \left\{
Li_3(e^{z^1-z^2}) + Li_3(e^{z^2-z^1})\right\}
\label{Fstar}
\end{equation}
where $Li_3(x)$ is the trilogarithm function and
\begin{eqnarray*}
t^1 & = & -\left(e^{z^1}+e^{z^2}\right)\,,\\
t^2 & = & z^1+z^2\,,
\end{eqnarray*}
and
\[
{\hat F}^\star = \frac{1}{4} \Big\{
({\hat z}^1)^2 \log ({\hat z}^1)^2+({\hat z}^2)^2 \log ({\hat z}^2)^2-
({\hat z}^1-{\hat z}^2)^2 \log({\hat z}^1-{\hat z}^2)^2
\Big\}
\]
where
\begin{eqnarray*}
{\hat t}^1 & = & {\hat z}^1 \, {\hat z}^2 \,, \\
{\hat t}^2 & = & {\hat z}^1 + {\hat z}^2\,.
\end{eqnarray*}

\end{example}

The significance of these dual coordinates will become apparent later. Note that with a further
change $z^1=w^2-iw^1\,,z^2=-w^2-iw^1\,,$ the dual prepotential (\ref{Fstar}) takes the form
\[
F^\star =\frac{1}{2} \sum_{k=\pm 1} Li_3(e^{2kw^2})+ 2i\left[\frac{1}{6} (w^1)^3  + \frac{1}{2} w^1(w^2)^2\right]
\]
and so falls into the class of solutions of the WDVV equations studied in \cite{MH}\,.

Schematically we have the following structure:
\[
\begin{array}{ccc}
F&\overset{S_\kappa}{\longrightarrow}&
{\hat F}\\
\downarrow & & \downarrow \\
F^\star
&&
{\hat F}^\star
\end{array}
\]
The aim of this note is two-fold: firstly to construct a transformation
\[
F^\star \overset{{\hat S}_\kappa}{\longrightarrow} {\hat F}^\star\,.
\]
This will turn out to be a twisted-Legendre transformation, where the twist is provided by the Euler vector
field. Secondly, the above example will be generalised to arbitrary dimension. This provides a
transformation between certain rational and
trigonometric solutions of the WDVV equations.

\section{Twisted Legendre Transformations}

We summarize the various structure in the following diagram:

\[
\begin{array}{ccc}
\left\{
<a,b>\,,\circ\,,E
\right\}&
\overset{S_\kappa}{\longrightarrow}&
\left\{
<a,b>_\kappa:=<\partial_\kappa\circ\partial_\kappa,a\circ b>\,,\circ\,,E
\right\}\\
& & \\
\downarrow & & \downarrow \\
& & \\
\left\{
\begin{array}{ccc}
(a,b) &:=& <E^{-1}\circ a,\,b>\\
a\star b &:=& E^{-1} \circ a \circ b
\end{array}
\right\}&&
\left\{
\begin{array}{ccc}
(a,b)_\kappa &:=& <E^{-1}\circ a,\,b>_\kappa\\
a\star b &:=& E^{-1} \circ a \circ b
\end{array}
\right\}
\end{array}
\]

\begin{prop}\label{twisted} There exists a vector field ${\hat\partial}_\kappa$
generating a twisted Legendre transformation
\[
F^\star \overset{{\hat S}_\kappa}{\longrightarrow} {\hat F}^\star
\]
such that
\[
(a,b)_\kappa = ({\hat\partial}_\kappa \star {\hat\partial}_\kappa,a \star b)\,.
\]
Explicitly,
\[
{\hat\partial}_\kappa = E \circ \partial_\kappa\,.
\]
\end{prop}

\begin{proof} The proof is straightforward:
\begin{eqnarray*}
(a,b)_\kappa & = & <E^{-1}\circ a,b>_\kappa\,,\\
& = & <\partial_\kappa\circ\partial_\kappa,E^{-1}\circ a \circ b>\,,\\
& = & (E\circ\partial_\kappa\circ\partial_\kappa,a \star b)\,,\\
& = & \left( (E \circ\partial_\kappa) \star (E \circ\partial_\kappa), a \star b\right)\,,\\
& = & ({\hat\partial}_\kappa \star {\hat\partial}_\kappa,a \star b)\,,
\end{eqnarray*}
on defining ${\hat\partial}_\kappa = E \circ \partial_\kappa\,.$

\end{proof}

It also follows from this result that if $\partial_{{\hat p}^a}$ are flat vector
fields for the metric $(,)_\kappa$ then ${\hat\partial}_\kappa \star\partial_{{\hat p}^a}$
are flat vector fields for the metric $(,)\,.$

\section{Hurwitz spaces and Frobenius structures}

Hurwitz spaces are moduli spaces of pairs
$(\mathcal{C},\lambda)\,,$ where $\mathcal{C}$ is a Riemann
surface of degree $g$ and $\lambda$ is a meromorphic function on
$\mathcal{C}$ of degree $N\,.$ It was shown in \cite{dubrovin1}
that such spaces may be endowed with the structure of a Frobenius
manifold. The $g=0$ case is particularly simple - meromorphic
functions from the Riemann sphere to itself are just given by
rational functions. It is into this category of Frobenius
manifolds that the examples constructed above fall.

More specifically, the Hurwitz space $H_{g,N}(k_1\,,\ldots\,,k_l)$
is the space of equivalence classes
$[\lambda:\mathcal{C}\rightarrow\mathbb{P}^1]$ of $N$-fold branched covers\footnote{Dubrovin uses the
different notation $H_{g,k_1-1\,,\ldots\,,k_l-1}\,.$}
with:

\begin{itemize}
\item $M$ simple ramification points $P_1, \dots,
P_M\in{\mathcal{L}}$ with distinct {\it finite} images $l_1,
\dots, l_M\in {\mathbb C}\subset {\mathbb P}^1$; \item the
preimage $\lambda^{-1}(\infty)$ consists of $l$ points:
$\lambda^{-1}(\infty)=\{\infty_1, \dots,\infty_l\}$, and the
ramification index of the map $p$ at the point $\infty_j$ is $k_j$
($1\leq k_j\leq N$).
\end{itemize}

\noindent  The Riemann-Hurwitz formula
implies that the dimension of this space is $M=2g+l+N-2\,.$ One has
also the equality $k_1+\dots +k_l=N$. For $g>0$ one has to introduce
a covering space, but this is unnecessary in the $g=0$ case that will be
considered here.

In this construction there is a certain ambiguity; one has to choose a so-called primary differential (also known
as a primitive form). Different choices produce different solutions to the WDVV equations, but such solutions are
related by Legendre transformation $S_\kappa\,.$ The Hurwitz data $\{\lambda,\omega\}$ from which one constructs a
solution $F_{\{\lambda,\omega\}}$ consists of the map $\lambda$ (also known as the superpotential) and a particular
primary differential $\omega\,.$ Thus, again schematically, one has:
\[
F_{\{\lambda,\omega\}} \overset{S_\kappa}{\longleftrightarrow} {\hat F}_{\{\lambda,{\hat\omega}\}}
\]
(note the map $\lambda$ does not change, though it might undergo a coordinate transformation). The metrics
$<,>\,,(,)$ and multiplications $\circ\,,\star$ are determined by calculating certain residues at the critial
points of the map $\lambda\,.$

\begin{thm}\label{superpotential}

\begin{eqnarray*}
<\partial',\partial{''}>& = & - \sum \res_{d\lambda=0}
\frac{\partial'(\lambda(v) dv) \partial{''}(\lambda(v)
dv)}{d\lambda(v)} \,, \\
<\partial'\circ\partial{''},\partial{'''}> & = & - \sum
\res_{d\lambda=0} \frac{\partial'(\lambda(v) dv)
\partial{''}(\lambda(v)
dv)\partial{'''}(\lambda(v)
dv)}{d\lambda(v)} \,, \\
(\partial',\partial{''}) & = & - \sum \res_{d\lambda=0}
\frac{\partial'(\log\lambda(v) dv) \partial{''}(\log\lambda(v)
dv)}{d\log\lambda(v)} \,, \\
(\partial'\star\partial{''},\partial{'''}) & = & - \sum
\res_{d\lambda=0} \frac{\partial'(\log\lambda(v) dv)
\partial{''}(\log\lambda(v)
dv)\partial{'''}(\log\lambda(v) dv)}{d\log\lambda(v)} \,.
\end{eqnarray*}

\end{thm}

\medskip

\noindent The first three formulae appeared in \cite{dubrovin1}
while the last follows immediately from the results in
\cite{dubrovin2}.

Rather than applying these methods to the full $g=0$ Hurwitz space we consider the space
$H_{0,k+m}(k,m)\,.$ This coincides with the space\footnote{This space is also isomorphic to the orbit space
${\widetilde W}^{(k)}(A_{k+m-1})$ corresponding to a certain extended affine Weyl group \cite{DZ}. This interpretation
will not be used here, but it does provide a starting point for the extension of these ideas to other,
non-Hurwitz, Frobenius manifolds.} of trigonometric polynomials of bidegree $(k,m)\,.$ Here the primary
differential is $dp$ and
\[
\lambda(p)=e^{ikp}+a_1 e^{i(k-1)p}+\ldots+a_k+\ldots a_{k+m} e^{-imp}\,.
\]
The coordinates $\{a_i\}$ are not flat coordinates for the metric $<,>\,$ but these may be constructed
via various residue formulae \cite{DZ}\,. The flat coordinates for the intersection form are just the
zeros of $\lambda\,.$

\begin{example} The prepotentials $F$ and $\hat F$ in examples \ref{example1} and \ref{example2} may be derived from the Hurwitz data
\[
\left\{\lambda(p) = e^p + t^1 + e^{t^2} \, e^{-p} \,, dp \right\}
\]
(here $\{t^1,t^2\}$ are the flat coordinates for the metric
$<,>$). The dual coordinates $\{z^1,z^2\}$ are the zeros of the
superpotential
\[
\lambda(p) = e^{-p} ( e^p - e^{z^1})( e^p - e^{z^2})\,.
\]
\end{example}

\noindent Thus in general one may write
\[
\lambda(p)=e^{-imp} \prod_{\alpha=1}^{k+m} \left(e^{ip} - e^{i z^\alpha}\right)\,.
\]
Using the above residue formulae one may construct the dual structure functions for the multiplication $\star$
and then integrate to find the dual prepotential ${F}^\star$ \cite{riley}\,.

\begin{prop}\label{resultA} The dual prepotential on the space of trigonometric polynomials of bidegree $(k,m)$ is:
\[
F^\star=A\sum_{i=1}^{k+m} (z^i)^3+
B\sum_{i=1}^{k+m} (z^i)^2 \sum_{j\neq j} z^j+ C\sum_{\overset{i,j,k}{distinct}} z^i z^j z^k +\frac{1}{2} \sum_{i\neq j} Li_3\left[e^{i(z^i-z^j)}\right]
\]
where
\begin{eqnarray*}
A&=&
\frac{i}{12} \left[ (m-2)(m-1)-mk\right]\,,\\
B&=&\frac{i}{4m}\left(2-m\right)\,,\\
C&=&\frac{i}{m}
\end{eqnarray*}
\end{prop}

\bigskip

An alternative way to construct the Frobenius structure on the Hurwitz spaces $H_{0,k+m}(k,m)$
is to use the superpotential (this is not quite true if $k_1=1$ but a similar form still holds)
\[
\lambda({\hat p})= {\hat p}^{k} + \sum_{r=0}^{k-2} a_r {\hat p}^r -
\sum_{\alpha=1}^{m}
\frac{c_{\alpha}}{({\hat p}-{\hat z}^0)^{\alpha}}\,,
\]
Geometrically this just corresponds to a change in primary differential and hence the prepotentials
are related via a Legendre transform.

\begin{example}

\noindent The prepotentials $\hat F$  and ${\hat F}^\star$ in examples \ref{example1} and \ref{example2} may be derived from the Hurwitz data
\[
\left\{ \lambda({\hat p}) = {\hat p} + \frac{{\hat t}^1}{{\hat p}-{\hat t}^2}\,, d{\hat p}\right\}\,.
\]
The dual coordinates $\{{\hat z}^1,{\hat z}^2\}$ are just the zeros of the superpotential
\[
\lambda({\hat p}) = \frac{ ({\hat p} - {\hat z}^1)({\hat p} - {\hat z}^2)}{({\hat p} - ({\hat z}^1+{\hat z}^2))}\,.
\]
On changing the primary differential:
\[
dp = \partial_{{\hat t}^1}[ \lambda({\hat p}) d {\hat p}]
\]
so $e^p = {\hat p}-{\hat t}^2$ the Hurwitz data transforms to
\[
\left\{
\lambda(p) = e^p + t^1 + e^{t^2} \, e^{-p} \,, dp \right\}
\]
and this generates the prepotential $F\,.$

\end{example}

As in the previous case, by factorizing $\lambda({\hat p})$ one may derive the dual prepotential very
simply. Explicitly one may write
\[
\lambda({\hat p}) = \left.
\frac{ \prod_{i=1}^{k+m} \left({\hat p}-{\hat z}^i \right)}{\left( {\hat p} - {\hat z}^0\right)^m}
\right\vert_{m {\hat z}^0 = \sum_{i=1}^{k+m} {\hat z}^i}\,.
\]
By calculating various residues and integrating to find the dual prepotential one finds:

\begin{prop}\label{resultBB} The dual prepotential on the Hurwitz space $H_{0,k+m}(k,m)$:
\[
{\hat F}^\star=\left.\frac{1}{4} \sum_{i,j=0\,,i\neq j}^{k+m} \alpha_i \alpha_j \left({\hat z}^i-{\hat z}^j\right)^2
\log\left({\hat z}^i-{\hat z}^j\right)^2\right\vert_{m {\hat z}^0 = \sum_{i=1}^{k+m} {\hat z}^i}\,,
\]
where $\alpha_0=-1$ and $\alpha_{i>0}=+1\,.$
\end{prop}

\noindent This part of the calculation may be generalised very simply to an arbitrary $g=0$ Hurwitz
space, and more generally to the induced structures on the discriminants of such Hurwitz space {\cite{FV,riley}}.
Since a change in primary differential induces a Legendre transformation between the corresponding
solutions of the WDVV equations, applying proposition \ref{twisted} yields the following result:

\begin{thm} The prepotentials $\hat F$ and ${\hat F}^\star$ constructed in propositions \ref{resultA} and \ref{resultBB}
are connected via a twisted Legendre transformation.
\end{thm}

\noindent This provides a map between certain rational solutions and trigonometric solutions to the WDVV
equations.

\section{Comments}

It would be interesting to extend these results to other classes of Frobenius manifold; all the
Hurwitz space examples consider above have an underlying $A_n$-structure but the construction
will work for arbitrary Weyl group. Such proofs would be more algebraic in style, using properties of the
roots systems (or, conjecturally, of $\vee$-systems \cite{sasha} via their geometrical interpretation in terms of
induced structures on discriminants \cite{FV}). Such \lq dual\rq~solutions are solutions of the WDVV equations
even though there is no \lq undual\rq-prepotential (the metric on the discriminant induced by $<,>$ has curvature,
though Frobenius-type structures remain \cite{iabs}). It would also be of interest to apply Legendre transformations
both twisted and non-twisted) directly to such solutions.

Also, given the close similarities between Calogero-Moser type operators and the flatness of the (dual) Dubrovin
connection, it would be of interest to see if the notion of a (twisted) Legendre transformation
holds for such operators, and in particular between rational and trigonometric operators.

\section*{Acknowledgments} Andrew Riley would like to thank the EPSRC for a research studentship.


\begin{thebibliography}{99}



\bibitem{dubrovin1} {Dubrovin, B., {\sl Geometry of 2D topological field theories} in {\sl Integrable
Systems and Quantum Groups}, ed. Francaviglia, M. and Greco, S.. Springer lecture
notes in mathematics, {\bf 1620}, 120-348.}

\bibitem{dubrovin2} Dubrovin, B., {\em On almost duality for Frobenius manifolds} in
{\sl Geometry, topology, and mathematical physics}, 75--132,
Amer. Math. Soc. Transl. Ser. 2, 212, Providence, RI, 2004.

\bibitem{DZ} Dubrovin, B. and Zhang, Y., {\em Extended affine Weyl groups and Frobenius manifolds},
Compositio Mathematica {\bf 111} (1998), 167--219.

\bibitem{FV} Feigin, M.V. and Veselov, A.P., {\sl Coxeter discriminants and logarithmic Frobenius structures},
math-ph/0512095.

\bibitem{MH} Martini, R., and Hoevenaars, L.K., {\sl Trigonometric Solutions of the WDVV Equations from Root Systems},
Lett.Math.Phys. {\bf 65} (2003) 15-18.

\bibitem{riley} {Riley, A.,} {University of Hull PhD thesis, 2006, in preparation.}

\bibitem{iabs} Strachan, I.A.B., {\sl Frobenius manifolds: natural submanifolds and induced bi-Hamiltonian
structures}, Differential Geom. Appl. {\bf 20} (2004) 67-99.

\bibitem{sasha} Veselov, A.P., {\sl Deformations of root systems and new solutions to generalised WDVV equations},
Phys. Lett. {\bf A 261} (1999), 297-302.


\end{thebibliography}
\end{document}